%% file: dyudina_apj_resubmission.tex
\documentclass[12pt,preprint]{aastex}

\usepackage{color}

\slugcomment{
See current version at http://arxiv.org/abs/1511.04415  

Revised after ApJ review.

Resubmitted on February 22, 2016

{\bf Revisions are shown in bold font according to ApJ instructions.}

Manuscript pages : \pageref{'last page'}; 

Submitted to ApJ on November 10, 2015 
}
\shorttitle{LIght curves and spherical albedos of Jupiter, Saturn, and exoplanets}
\shortauthors{Dyudina et al.}

\input{def_no_journals}

\def\cora#1{\textbf{\textcolor{black}{\rm#1}}}   
\def\corb#1{\textbf{\textcolor{black}{\bf#1}}}   
\def\corc#1{\textbf{\textcolor{black}{\rm#1}}}
\def\cord#1{\textbf{\textcolor{black}{\rm#1}}}

\begin{document}
\bibliographystyle{icarus}

\title{\cora{Reflected Light Curves, Spherical and Bond Albedos of Jupiter- and Saturn-like Exoplanets}}
\author{Ulyana Dyudina\altaffilmark{1}, Xi Zhang\altaffilmark{2},  Liming Li\altaffilmark{3}, 
 Pushkar Kopparla\altaffilmark{1}, Andrew P. Ingersoll\altaffilmark{1}, Luke Dones\altaffilmark{4}, Anne Verbiscer\altaffilmark{5}, and Yuk L. Yung\altaffilmark{1}}
\affil{(1)Division of Geological and Planetary Sciences, 
 150-21 California Institute of Technology,  Pasadena, CA 91125 (U.S.A.)
(2) University of California Santa Cruz 1156 High St, Santa Cruz, CA 95064, (U.S.A.)
(3) Department of Physics, University of Houston, Houston, TX 77204, (U.S.A)
(4) Southwest Research Institute, 1050 Walnut St., Suite 300, Boulder CO 80302 (U.S.A.)
(5) Department of Astronomy, University of Virginia, Charlottesville, VA 22904-4325 (U.S.A.)}




\altaffiltext{1}{Division of Geological and Planetary Sciences, 
 150-21 California Institute of Technology,  Pasadena, CA 91125 (U.S.A.), ulyana@gps.caltech.edu}

\include{abstr_include}


\input{intro_include}

\input{model_include}

\input{results_include}

\input{discussion_include}

\include{supplementary_captions}

\section*{Acknowledgements}
This research was supported by the NASA Cassini Project.
We thank R.A. West for his critical comments on the manuscript, references on Jupiter's scattering, and help with image calibration.
We thank S.P. Ewald for the comments on the manuscript.  
We thank the anonymous reviewer for useful suggestions. 
\label{'last page'}
\bibliography{my}
\end{document}

%% file: def_no_journals.tex
\def\mm{$\mu$m }
\def\mmdot{$\mu$m}
\def\ni{\noindent}  
\def\tms{$\times$}  
\def\ie{{ i.e.}}  
\def\eg{{ e.g.}}  
  
\def\deg{$^{\circ}$}
\def\nh3{ $\mbox{NH}_3$}
\def\ch4{$\mbox{CH}_4$}
\def\h2o{ $\mbox{H}_2\mbox{O}$}
\def\c2h2{ $\mbox{C}_2\mbox{H}_2$}


%% file: abstr_include.tex
\begin{abstract}

\corb{Reflected light curves observed for exoplanets indicate bright clouds at some of them.} 
We estimate how the light curve and total stellar heating of a planet depend on forward and backward scattering \corb{in the} clouds  
\corb{based on Pioneer and Cassini spacecraft images of Jupiter and Saturn}.
We fit analytical functions to the local reflected brightnesses of Jupiter and Saturn \corb{depending on the} planet's phase.
These observations cover broad bands at 0.59-0.72 and  0.39-0.5 \mmdot, and  narrow bands at 0.938 (atmospheric window), 0.889 (\ch4 absorption band), and 0.24-0.28 \mmdot.
We simulate the images of the planets 
with a ray-tracing model, 
\corb{ and disk-integrate them to produce} the full-orbit light curves.
For Jupiter, we 
\corb{ also fit the modeled light curves to the} observed full-disk brightness. 
We \corb{derive} 
spherical albedos for Jupiter, Saturn, and for planets with Lambertian and Rayleigh-scattering atmospheres.
Jupiter-like atmospheres \cora{can} produce light curves that are \cora{a factor of two} fainter at half-phase than the Lambertian planet, given the same \cora{geometric albedo} at transit.
The spherical albedo 
is \corb{typically} lower than for a Lambertian planet \corb{by up to a factor of $\sim$1.5}.
\corb{The Lambertian assumption will 
underestimate the absorption of the stellar light and the equilibrium temperature of the planetary atmosphere.}
\corb{We also compare our light curves with the light curves of solid bodies: the moons Enceladus and Callisto. 
Their strong backscattering peak within a few degrees of opposition (secondary eclipse) can lead to an even stronger underestimate of the stellar heating.}

\end{abstract}

%% file: intro_include.tex
\section{Introduction}
\label{sec: introduction}

In the last decade thermal light curves and secondary eclipses were observed for more than 50 transiting exoplanets, and \corb{several} were directly imaged \citep{madhusudhan14}.
Most of these observations are either time-varying signals (transits, secondary eclipses, and phase curves), or spatially resolved imaging and spectroscopy. 
These techniques allow \cora{one} to study the relative abundances of common elements such as C, H, O, and N, corresponding atmospheric chemistries, vertical pressure-temperature profiles, global circulation patterns, and clouds, which block the emission from the planets and screen the planet from the star's heating.

The first phase curves of exoplanets showed thermal emission.
Starting with the non-transiting planets {$\upsilon$} Andromedae b at 24\mm \citep{harrington06} and HD 179949 at 8\mm\citep{cowan07}, and the transiting hot Jupiter HD 189733b at 8\mm \citep{knutson07},  phase curves have been detected for about a dozen planets. 
Visible-wavelength \corb{full-orbit phase curves were detected and modeled for a few planets \citep{quintana13, esteves13, esteves15, shporer15}}.
\corb{In} visible wavelengths the phase curve is the sum of the thermal and reflected light components, and infrared observations are required to distinguish between the two.
\cite{demory13} identified the reflected nature of the phase curve of the hot Jupiter Kepler-7b  by comparing it to the 3.6 and 4.5\mm $Spitzer$ observations, showing that the planet is highly reflective.
Its geometric albedo (the ratio of the planet's reflected flux to that of a same-size flat Lambertian disk) is $A_g=0.35\pm0.02$.
Another detection of a blue-colored reflective ($A_g=0.4\pm0.12$) hot Jupiter HD 189733b comes from the secondary eclipse \citep{evans13}.
High reflectivity suggests clouds, which may originate from condensation of silicates or other ``rocky'' materials, or from photochemical \corb{and ion-chemical} processes.
Most hot Jupiters have \corb{geometric} albedos less than 0.1, but a subset have much larger albedos around 0.3 \citep{heng13}.
\corb{No multi-wavelength reflected light curves have been detected yet.}  

\corb{Wavelength-dependent phase functions and albedos have been modeled for different distances from the star with cloud condensation and light scattering microphysics models \citep{marley99, burrows04, sudarsky05}.
Different atmospheric compositions and radiative -- convective equilibrium models were tested to study the effects on the light curves and spectra \citep{cahoy10}.
\cite{kane10} modeled 550-nm light curves for multi-planet systems on long-period eccentric orbits.
Different models assume Lambertian surface scattering, Rayleigh-scattering atmospheres \citep{madhusudhan12}, or surface scattering produced by modeled microphysics \citep{burrows04, sudarsky05} or observed by spacecraft at  Jupiter and Saturn \citep{dyudina05}.
}

\corb{The light curves of Jupiter and Saturn can provide guidance on possible shapes of light curves of hotter, more detectable exoplanets.}
Reflected light from Jupiter and Saturn analogs is \corb{faint and hard to detect} 
(the relative planet to star flux ratio F/F$_ *\sim 10^{-8}-10^{-9}$, or 0.001-0.01 ppm).
The reflected light from hot Jupiters is detected at the levels F/F$_ *\sim 10^{-4}-10^{-5}$. 
The detections indicate albedos up to $A_g\sim0.4$, which requires clouds.
Though clouds on Jupiter and Saturn form at different temperatures, larger pressures, from different chemical elements, and are somewhat brighter ($A_g\sim0.5-0.6$), they represent multiple scattering on relatively bright cloud particles, as on hot Jupiters.
The spectrum of the clouds on Jupiter and Saturn, sampled in atmospheric windows, is surprisingly featureless.
\corb{Even with the vast observations available, theoretically predicted water and ammonia cloud compositions \citep{weidenschilling73} were not observationally confirmed until the late 1990-s \citep{saturn10b_clouds}.
Visible spectra of Jupiter and Saturn are dominated by \ch4 absorption bands which originate from the atmosphere above the clouds.}
These clouds act as spectrally flat scatterers, whose brightness depends on the particle size, \corb{shape, single scattering} albedo, and \corb{number} density, but not on composition.
Accordingly, clouds on extrasolar planets may have similar scattering properties in atmospheric windows, even though different gases above the clouds can result in different spectra.
Models of cloud \corb{coverage} on Jupiter and Saturn from observations 
give a wide range of possible cloud properties \citep[reviewed by][]{west04, saturn10b_clouds}.
Application of cloud models to a disk-integrated planet's luminosity, which is needed for orbital light curves, inherits the uncertainty.

This paper is the first to use the Cassini Jupiter flyby visible images to construct Jupiter's light curve \corb{for various wavelengths}.
This data set \corb{from 2000-2001 \citep{porco03,zhang13}} has the best phase angle coverage among spacecraft observations. 
We also use published data from previous spacecraft that visited Jupiter and Saturn.
The most complete published data on the \cora{reflectance} of  Jupiter's and Saturn's cloud top \cora{``surfaces"} come from Pioneer 10 and 11 \citep{tomasko78,smith84,tomasko84}.
The \cora{reflectances} were measured for two \corb{broad} wavelength bands (red and blue). 
The \cora{phase angle dependence of \corb{these surface} reflectances} was used to reconstruct light curves of a planet similar to Jupiter and Saturn \cora{using} a 3D model of a planet with or without rings \citep{dyudina05}.
Jupiter's and Saturn's atmospheres show strong backward and forward scattering. 
As a result the light curve shapes were substantially different from the Lambertian case. 

We also derive a planet's spherical albedo \corb{for} various wavelength bands - the reflective property, closely related to Bond albedo, which  characterizes the total reflected flux from the planet.
Bond albedo for extrasolar planets had been  estimated from their visible luminocities at secondary eclipse \corb{(\ie, from their geometric albedos $A_g$)} and \corb{by balancing} the flux radiated in the infrared \citep{schwartz15}.
The poorly restricted \corb{Bond-to-geometric albedo ratio is explored} in this paper using Jupiter and Saturn as examples.
\corb{This ratio had been previously addressed  theoretically for plane-parallel semi-infinite atmospheres \citep{van_de_hulst80b}, including fits to $A_g$ and to the center of disk reflectivities of Saturn and Uranus \citep{dlugach74}.
\cite{hovenier89} modeled the Bond albedo using single-scattering cloud particle phase functions derived from observations of Venus \citep{whitehill73, hilton92} and from Pioneer observations of Saturn \citep{tomasko84}.}

\corb{Cassini observations provided unprecedented wavelength and phase angle coverage for constructing light curves and spherical albedos for different wavelengths.}
We derive \cora{light curves} from Cassini Jupiter flyby images.
We also use surface reflectance phase functions derived by \cite{dyudina05} from Pioneer data \citep{tomasko78,smith84}.

The planet model and the planets' \cora{measured reflectances} are described in Section \ref{sec:model}.
The light curves \cora{for different wavelength bands} are \corb{modeled} in Section \ref{sec:jupiter_saturn}.
The planet-integrated reflected light (spherical albedos) for Jupiter and Saturn at different wavelength bands, and the range of possible stellar heating for extrasolar cloud-covered planets are discussed in Section \ref{sec: spherical_albedo}.
Possible implications of our results for the extrasolar planets are discussed in Section \ref{sec: discussion}.
\corb{The digital version of the light curves derived in this work, as well as the digital Cassini data used to derive them, are given as an online supplement.}

%% file: model_include.tex
\section{Model}
\label{sec:model}

We \cora{simulate planet images} by tracing plane-parallel light rays from the distant central star reflected by each position on the planet \citep{dyudina05}.
This produces images (80 pixels across the planet's disk).
In this work the planet is spherical.
The wavelengths in our model are integrated across the transmissivity of the Pioneer \corb{blue (0.39-0.5 \mmdot) and red (0.59-0.72 \mmdot)  filters and Cassini UV1 (0.24-0.28 \mmdot), MT3 (0.889 \mmdot), and CB3 (0.938 \mmdot) filters, as will be discussed later}.
Our  notation matches that of most observational papers on Saturn and Jupiter 
\corb{ (see \cite{dyudina05})}. 
The variables we use in this paper are defined in Table \ref{tab:def}.
\noindent
\begin{table}[htbp]
\begin{small}
\begin{tabular}{|c|c|c|c|c|c|c|c|}
\hline
\noindent
Variable      & Description & Units (if any) \\
\hline
\noindent
$A,B$          &Coefficients of the Barkstrom law \cora{(Eq. \ref{eq: backstorm})}              & \\
$A_b$          &Bond albedo             & \\
$A_g$          &Geometric albedo             & \\
$A_{HG}$	&Coefficient of Henyey-Greenstein function             &\\
$A_S$          &Spherical albedo             & \\
$F$   &Intensity of a white Lambertian surface\tablenotemark{a}
                                                                    &$Wm^{-2}sr^{-1}$  \\
$g_1,g_2,f$ &Parameters of double Henyey-Greenstein function          & \\
$I$    &Intensity (or brightness, or radiance) of the surface&$Wm^{-2}sr^{-1}$\\
$L_P$    &Luminosity of the planet\tablenotemark{b}         &$W sr^{-1}$\\
$L_*$   &Luminosity of the star\tablenotemark{b}        &$W sr^{-1}$\\
$p(\alpha)$        &Full-disk  albedo\tablenotemark{b}\tablenotemark{c}   $L_P/(\pi R_P^2 F)$ &\\
$R_P$   & Radius of the planet                   &km\\
$r_{\rm pix}$ &Pixel size                                        &km\\
$\alpha$&Phase angle                                       &degrees\\
$\Theta$&Orbital angle ($\pm$180\deg: min phase, 0\deg: max phase)&degrees\\
$\mu_0,\mu$&Cosines of the incidence and emission angle&\\
\hline     
\end{tabular}
\caption{Variables used in our modeling.
The detailed definitions follow in the text.}
\label{tab:def}
\end{small}
\tablenotetext{a}{$F\cdot(\pi \rm{~ steradians})$ is the incident stellar flux at the planet's orbital distance (which is also sometimes called $F$, but has $W m^{-2}$ units, unlike our intensity $F$ measured per unit solid angle).}
\tablenotetext{b}{The optical properties for particular filters are the convolution of the planet's spectrum with the wavelength-dependent filter transmissivity.}
\tablenotetext{c}{The ``full disk albedo" \cora{term is adopted from \cite{karkoschka98}.
It is} related to the variable $\Psi$ used by \cite{seager10b}: $\Psi=\pi \cdot p(\alpha)$.}
\end{table}

\subsection{Reflecting properties of Jupiter, Saturn, Lambertian and Rayleigh-scattering planets}

We study the light curves and total solar absorbed light \cora{as they depend on scattering by the planet's cloudy surface}.
We do not distinguish the altitudes at which clouds scatter the light, but use the observed planet's brightness, representing light scattered by an entire atmospheric column.
We test \cora{a set of analytical functions describing surface scattering} of Jupiter and Saturn \corb{by visually fitting them} to spacecraft observations.
\corb{We} compare them to a \corb{(constant with phase angle $\alpha$) Lambertian scattering function}, and to modeled \corb{surface scattering of a} semi-infinite Rayleigh-scattering atmosphere 
\corb{(with particle single scattering albedos of 0.999999 and 0.3)} \citep{kopparla15}.
When possible, we use published data on the planet's surface scattering, \eg, the already fitted \cora{analytical function} for Saturn by \cite{dones93}.
\cora{For Jupiter these surface scattering measurements were \corb{used in the cloud model} by \citep{tomasko78} in order to derive the single scattering phase function of cloud particles.
We do not attempt to \corb{do the same and derive} the cloud \corb{distribution and single-particle scattering for} Pioneer observations.
Instead we \corb{summarize the reflectivity data published by \cite{tomasko78} with the help of} analytical functions  
\corb{and use these functions} to simulate full-disk images.}

\subsubsection{Jupiter}
\label{sec: jupiter}

\cora{Each location on Jupiter's surface is assumed to reflect solar light in the same way, \ie, our analytical reflectance function is uniform over the planet.
In general, the reflected light} depends on three angles: incidence (via the cosine of incidence angle $\mu_0$), emission (via the cosine of emission angle $\mu$), and phase angle $\alpha$.
In our notation, the phase angle $\alpha=0$\deg\ indicates backward scattering and $\alpha=180$\deg\ indicates forward scattering.
We ignore the dependence on $\mu$ and \cora{assume} brightness proportional to $\mu_0$ for Jupiter because, \cora{as we will show, this simplification can give a reasonably good fit to the data.}

We fit an analytical function to \cora{the image pixel brightness} in the units of $I/F$.
\begin{equation}
I(\alpha,\mu_0, \mu)/F={\mu_0}\cdot P( A_{HG}, g_1,g_2,f,\alpha).
\label{eq: iof}
\end{equation}
$I(\alpha,\mu,\mu_0)$ is the reflected intensity at a given location on the planet.
$\mu_0$ is the cosine of the incidence angle measured from the local vertical.  
$F\cdot\mu_0$ is the reflected brightness of a white Lambertian surface.
$F\cdot(\pi \rm{~ steradians})$ is the solar flux at the planet's orbital distance. 
$P( A_{HG}, g_1,g_2,f,\alpha)$ is a two-term Henyey-Greenstein function.
\begin{equation}
P( A_{HG}, g_1,g_2,f,\alpha)= A_{HG}\cdot(fP_{HG}(g_1,\alpha)+(1-f)P_{HG}(g_2,\alpha))
\label{eq:h_g}
\end{equation}
The coefficient $A_{HG}$ is fitted to match the amplitude of the observed phase function.
The individual terms $P_{HG}$ are Henyey-Greenstein functions representing forward and backward scattering lobes, \cora{respectively}.
 \begin{equation}
P_{HG}(g,\alpha)\equiv{{(1-g^2)}\over{(1+g^2+2g \cdot \cos{\alpha})^{3/2}}}~~,
\end{equation}
where $\alpha$ is the phase angle, \cord{$f\in[0,1]$} is the fraction of forward versus backward scattering, and $g$ is one 
of $g_1$ or $g_2$;  
\cord{$g_1\in[0,1]$} controls the sharpness of the forward scattering lobe, while \cord{$g_2\in[-1,0]$} controls the sharpness of the backscattering lobe.

\cora{A commonly-used expression for surface reflection, the Bidirectional Reflection Distribution Function (BRDF) can be expressed in \corb{terms} of Eq. \ref{eq: iof}:  BRDF = $I(\alpha,\mu_0, \mu)/(\pi F \mu_0)$.
Accordingly, our  double Henyey-Greenstein function approximation is a scaled BRDF: BRDF $\approx P( A_{HG}, g_1,g_2,f,\alpha)/\pi$.}
Note that typically the Henyey-Greenstein function is used for single-particle scattering, and then the function is normalized over the emission solid angles to give a unit single scattering albedo.
Here we use this function only as a convenient analytical expression to \corb{represent} the measured \corb{result of} multiple scattering of Jupiter's cloud surface, for which the particles' single scattering albedo is not relevant.
In our case, 
the value to normalize \corb{by} is the spherical  albedo \corb{$A_S$} (the ratio of reflected to incident light for the whole planet at appropriate wavelengths).  
It is imbedded in the Henyey-Greenstein function.
We will calculate the spherical albedos later in this paper. 

\begin{figure}[htbp]
\vspace{-.5cm}
 \begin{minipage}[c]{0.5\textwidth}
\resizebox{3.5in}{!}{\includegraphics{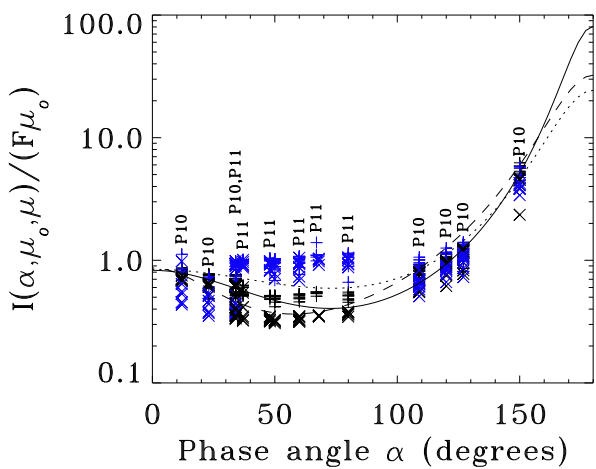}}
\vspace{-0.3in}
  \end{minipage}\hfill
  \begin{minipage}[c]{0.48\textwidth}
\caption{
\corc{Fits of Henyey-Greenstein functions to the Pioneer 10 data, labeled P10 \citep{tomasko78}, and Pioneer 11 data, labeled P11 \citep{smith84}. 
The data represent belts (\tms\ symbols) and zones (+ symbols) on Jupiter observed with the red (0.595--0.720 \mmdot, black symbols) and blue (0.390--0.500 \mmdot, blue symbols) filters.
\cora{The solid, dashed and dotted lines demonstrate the range of possible fits to the red filter points (black symbols).}}
}
\label{fig:tomasko78fit}
  \end{minipage}\hfill
\end{figure}

Figure \ref{fig:tomasko78fit} shows \cora{ our fits} of Henyey-Greenstein functions to the \cora{image pixel values} 
from the Pioneers 10 and 11 \cora{flybys} \citep{tomasko78,smith84}. 
The corresponding Henyey-Greenstein coefficients are given in Table  \ref{tab: hg}.
\begin{table}[htbp]
  \begin{center}
       \begin{tabular}{|c|c|c|c|c|c|c|c|c|c|c|}
\hline
\noindent
Spacecraft & Year & Filter & Wavelength  & Linestyle & $A_{HG}$ & $g_1$ & $g_2$ & $f$\\
\hline
Pioneers 10,11& 1973,1974&red& 0.595-0.72 \mm& solid&2&0.80&-0.38&0.90 \\
 & & & & dashed&1.80&0.70&-0.55&0.95 \\
 & & & & dotted&1.60&0.70&-0.25&0.80 \\
\hline
Cassini& 2000-2001&CB3& 0.938 \mm & solid&1.07&0.60&-0.30&0.80 \\
 & & & & dashed&1.20&0.70&-0.40&0.87 \\
 & & & & dotted&0.97&0.35&-0.10&0.65 \\
\hline
Cassini& 2000-2001&MT3& 0.889 \mm & solid&0.11&0.35&-0.35&0.93 \\
 & & & & dashed&0.11&0.30&-0.70&0.99 \\
 & & & & dotted&0.09&0.50&-0.10&0.60 \\
\hline
Cassini& 2000-2001&UV1& 0.258 \mm & solid&0.60&0.40&-0.40&0.80 \\
 & & & & dashed&0.40&0.50&-0.50&0.80 \\
 & & & & dotted&0.50&0.40&-0.20&0.60 \\
  \hline     
      \end{tabular}
    \caption{\cora{Henyey-Greenstain coefficients fitted to Pioneer and Cassini Jupiter data.}
    } 
    \label{tab: hg} 
  \end{center}
\end{table}
The solid line was used by \cite{dyudina05} to fit the black data points (red filter) in Fig. \ref{fig:tomasko78fit}.
In this paper we also show the dotted and dashed curves, which represent the range of functions 
that are visually consistent with the range of black data points in Fig. \ref{fig:tomasko78fit}.

Pioneer images show\cora{ that different surface locations on Jupiter have} different scattering properties.
In particular, \cite{tomasko78} and \cite{smith84} indicate two types of locations: the belts, usually seen as dark bands on Jupiter, and zones, usually seen as bright bands on Jupiter.
\corb{As discussed in \cite{dyudina05},} the relative calibration between Pioneer 10 and Pioneer 11 data is not as well constrained as the calibration within each data set.
Our model curve better represents the observations in the red filter (black data points) than in the blue filter.
The Pioneer 11 blue data points at moderate phase angles $\alpha\sim 30-80${\deg} (P11 in Fig. \ref{fig:tomasko78fit}) seem to be systematically offset up from the Pioneer 10 (P10) blue data points.
\corb{This} may be a result of relative calibration error, or temporal change in the clouds. 
Because of that, we do not model the blue wavelengths.
Red wavelengths may also have that problem, though it is not so obvious in Fig. \ref{fig:tomasko78fit} (black data points).
Accordingly, the red phase function derived by \cite{dyudina05} (black \cora{solid} curve in Fig. \ref{fig:tomasko78fit}) is also uncertain.
\cora{However}, after disk averaging, this fitted function reproduces the full-disk brightness observed by HST \citep{karkoschka94,karkoschka98} at the red passband of the Pioneer filters.
\cora{To explore the effects of the Pioneer 10/11 uncertainty in red wavelengths we fit two other curves (dotted and dashed) to the data.}



\begin{figure}[htbp]
\vspace{-.7cm}
\resizebox{3.in}{!}{\includegraphics{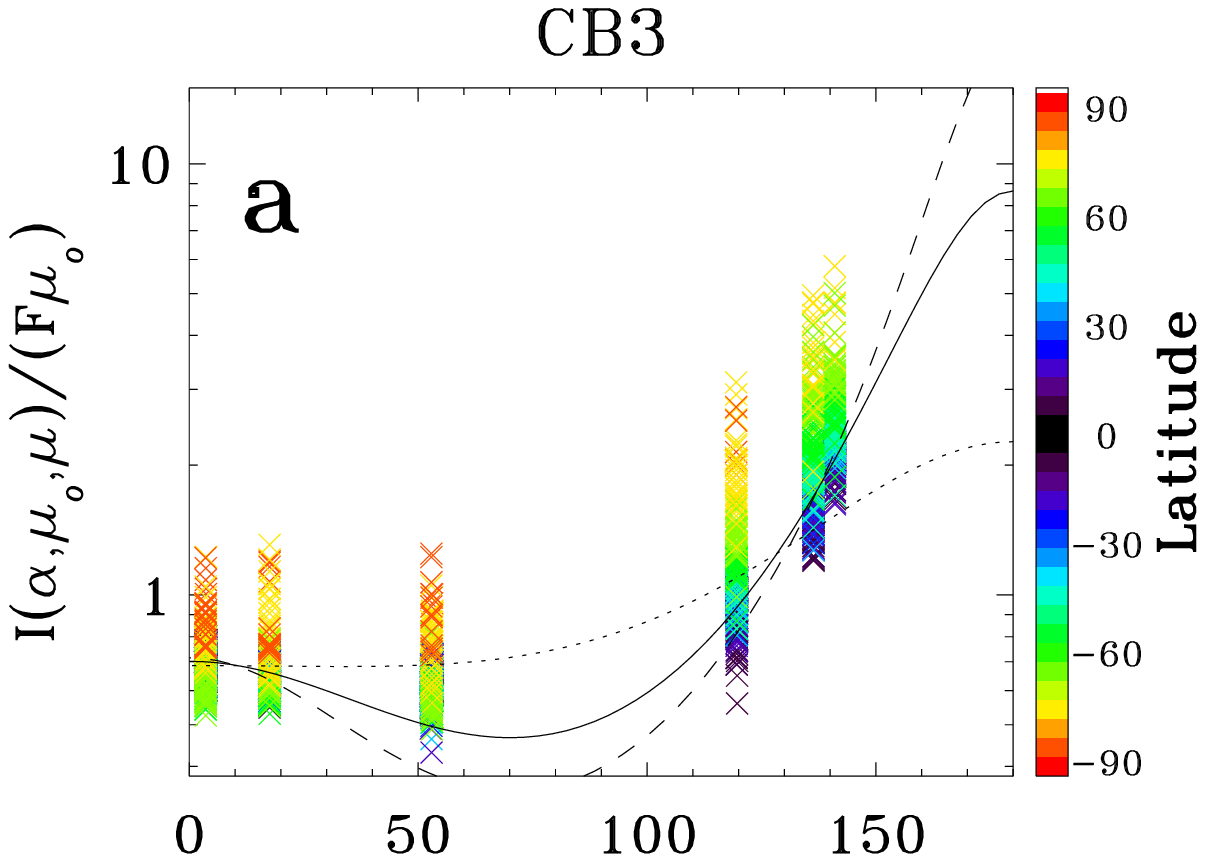}}
\hspace{1cm}
\resizebox{3in}{!}{\includegraphics{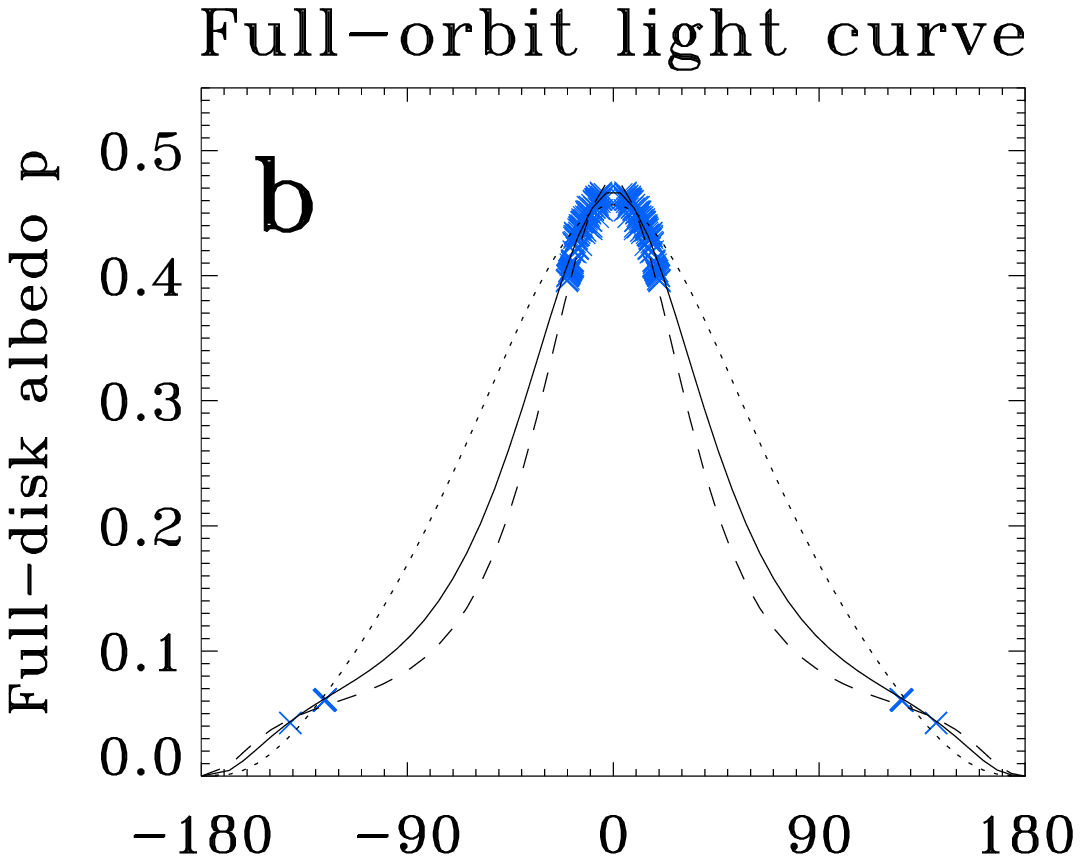}}
\vspace{-1.45cm}
\newline
\resizebox{3.in}{!}{\includegraphics{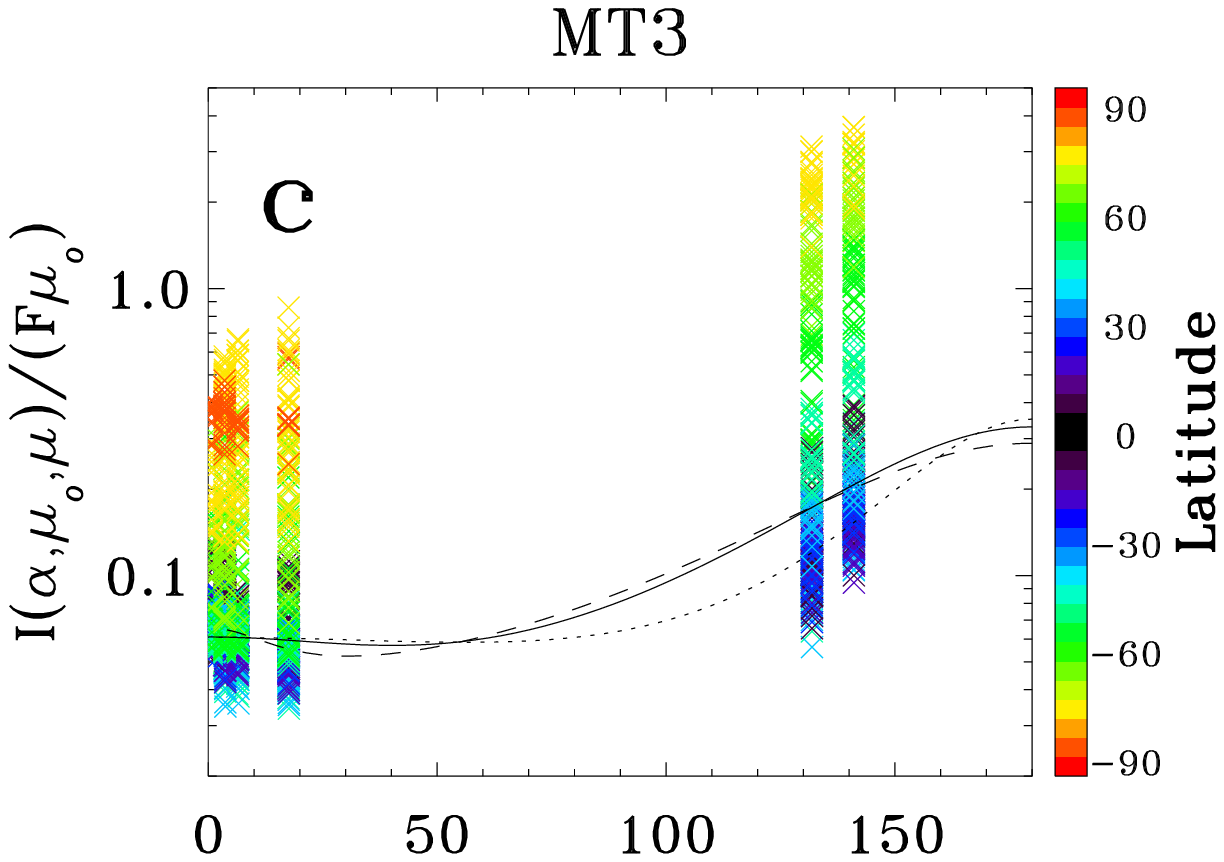}}
\hspace{1cm}
\resizebox{3in}{!}{\includegraphics{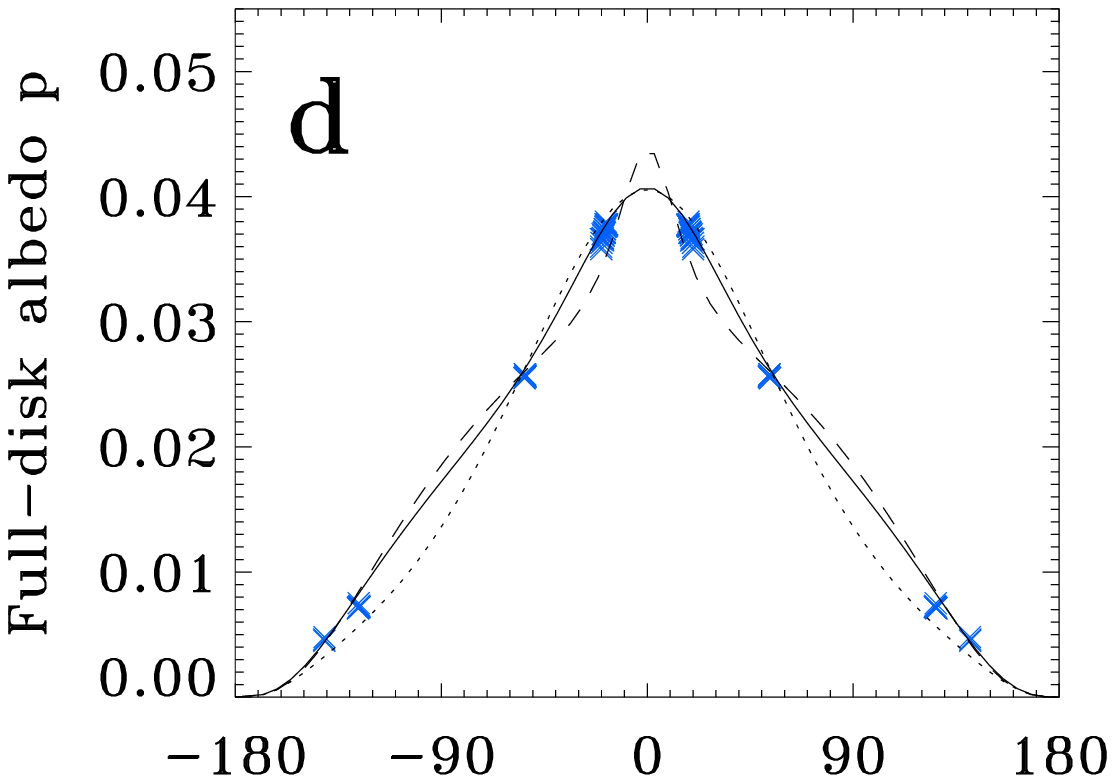}}
\vspace{-1.45cm}
\newline
\resizebox{3.in}{!}{\includegraphics{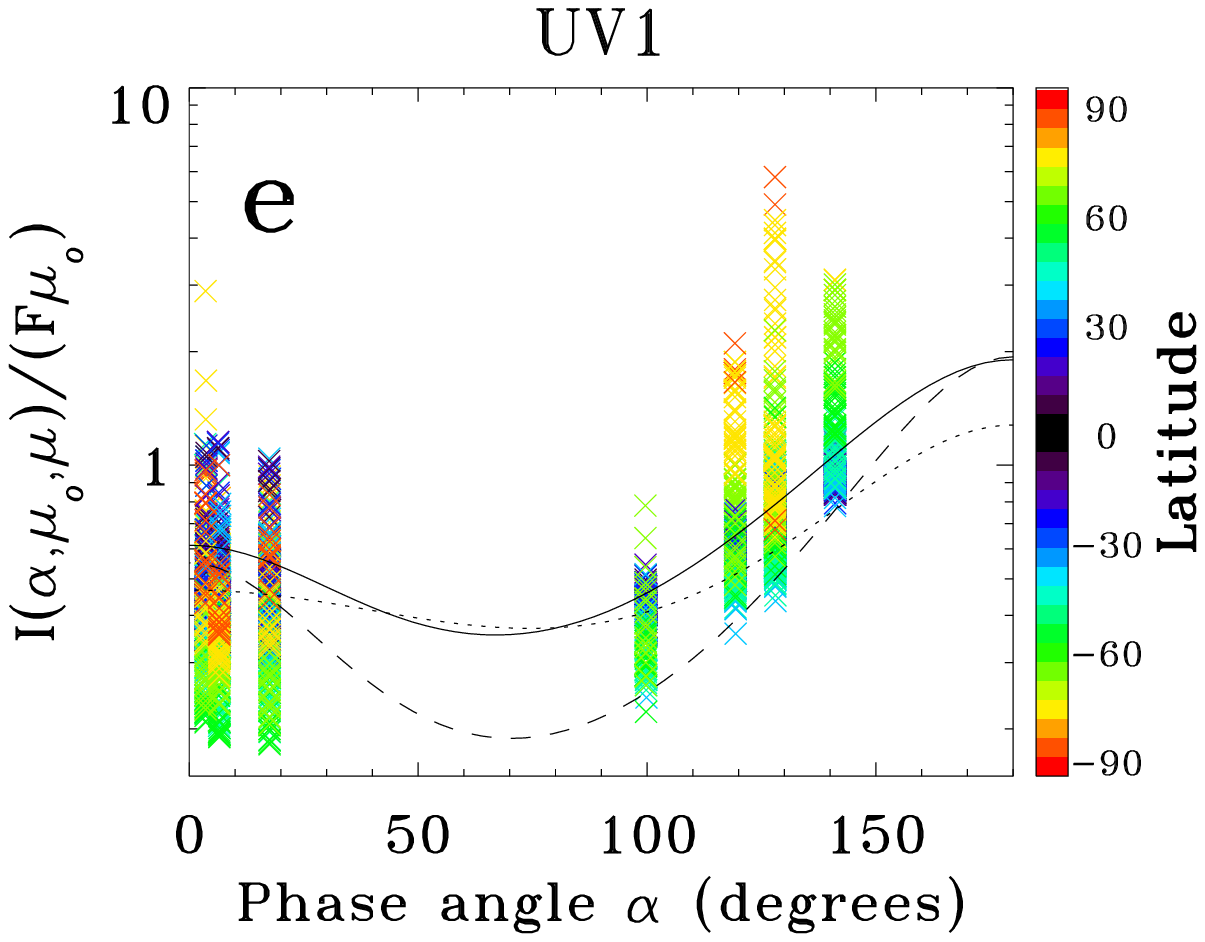}}
\hspace{1cm}
\resizebox{3in}{!}{\includegraphics{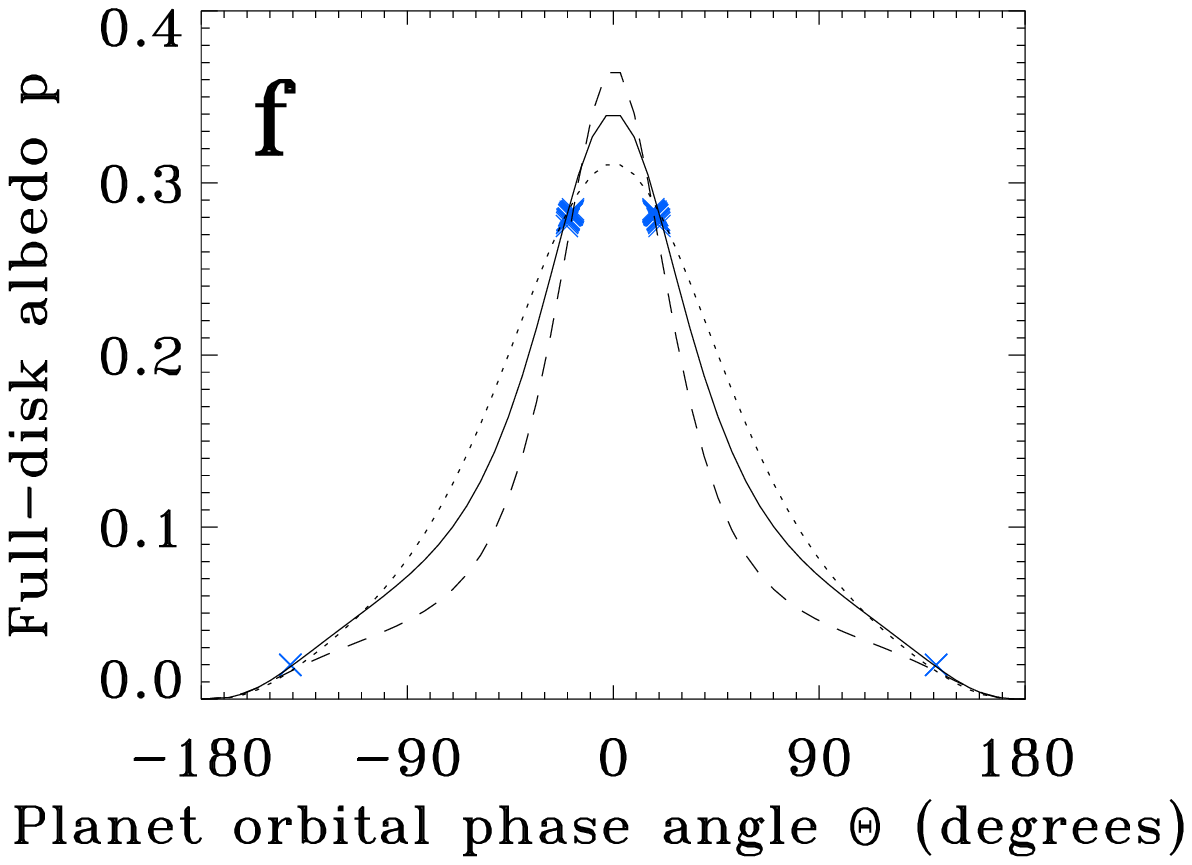}}
\vspace{-1.2cm}
\newline
\resizebox{3.in}{!}{\includegraphics{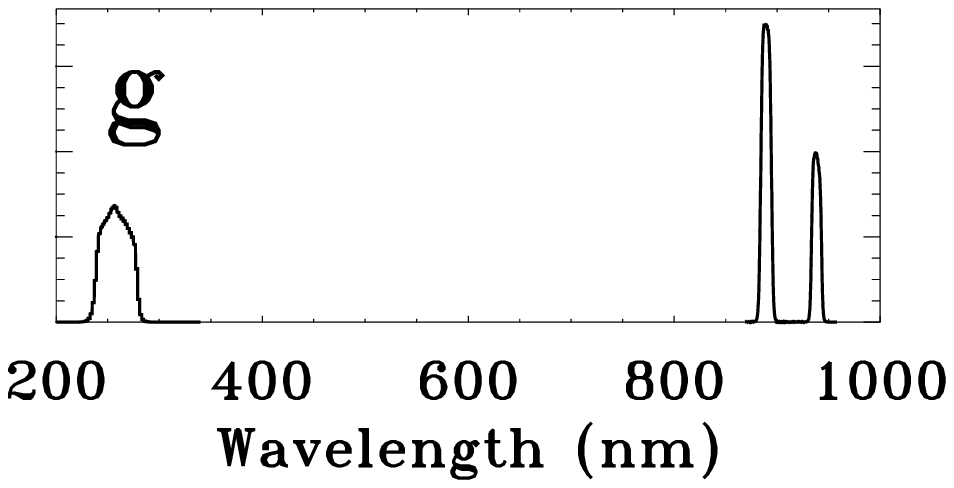}}
\vspace{-.7cm}
\caption{
\cora{Henyey-Greenstein functions (curves in panels a, c, e with linestyles corresponding to Table \ref{tab: hg}) fitted to \corb{image} pixel brightness} \corb{for the} Cassini 2000-2001 Jupiter flyby. 
Panels \cora{ a,c,e} represent images taken with \cora{Cassini NAC camera filters: CB3, MT3, and UV1, as labeled on top of the plots}.
The same data were used to model atmospheric aerosols by \cite{zhang13}.
\cora{The right panels (b, d, f) show the modeled light curves corresponding to the lines in the left panel for each filter.
The blue $\times$ symbols show disk-integrated brightness measured from the Cassini images.}
\corb{The digital data points for panels a-f are available as a supplement.}
Panel \cora{g}  shows filter shapes: UV1 at 258 nm, CB3 at 938 nm, and MT3 at 889 nm.
}
\label{fig: cassini_fit}
\end{figure}

Figure \ref{fig: cassini_fit} shows our fits of the  Henyey-Greenstein functions \cora{(panels a,c,e)} to the data from the Cassini images of Jupiter. 
The functions are fitted in the atmospheric window (938 nm CB3 filter), strong \ch4 absorption band (889 nm MT3 filter), and in the ultraviolet (258 nm UV1 filter).  
\corb{The atmospheric window band probes the deepest in the atmosphere, and is sensitive to the clouds at a variety of depths.
The strong \ch4 absorption band is sensitive only to the highest clouds and hazes.
The ultraviolet channel is sensitive to the upper haze layer and Rayleigh scattering.
The filter details can be found \corb{in Fig. \ref{fig: cassini_fit} g and in 
\cite{porco04,zhang13}}.
The data points are available in digital form as supplementary online material.}
The fitted Henyey-Greenstein parameters are listed in  Table  \ref{tab: hg}.
The \corb{image pixel brightness data points (panels a, c, e)} represent a variety of incidence and emission angles.
The high-latitude data points (orange
) are illuminated and observed at very slanted angles during the equatorial Cassini flyby.
These points are systematically \cora{higher due to limb brightening and our non-perfect accounting for illumination as 1/$\mu_0$}.
Because of that, we ignored the high-latitude data points and fitted the phase functions to the lower-latitude data points (green to blue in \cora{the left panels of} Fig. \ref{fig: cassini_fit}).
\cora{The Henyey-Greenstein curve fits for the local pixel brightness values in the left panels (input to our disk-averaging model) were also tuned such that the the disk-integrated brightness curves in the right panels (output from the model) fit the disk-averaged data points  measured from the Cassini full-disk images \corb{(blue $\times$ symbols)}.}
\cora{The full-disk images were taken during the October, 2000 -- March, 2001 flyby.
We use images with spatial resolutions ranging $\sim$200-2000 km/pixel to compute the full-disk albedo. 
The images are calibrated by the Cassini ISS CALibration software (CISSCAL)  {\footnotemark} \citep{west10}.
\footnotetext{
\corb{CISSCAL performs standard CCD calibration such as bias/dark subtraction,  flatfield correction, and ISS-specific calibrations. 
The ground-based observations \citep{karkoschka98} are used to improve this calibration. 
CISSCAL outputs the absolute reflected solar irradiance in units of photons/second/cm$^2$/nm/steradian for the wavelength bands of Cassini filters.} }
We compute the full-disk albedo as the disk-averaged $I/F$ 
for different Cassini filters{\footnotemark}.\footnotetext{ 
\cora{\corb{The reference solar spectral irradiance is combined using data from the Upper Atmosphere Research Satellite (UARS, 1991 to 2001, including Cassini flyby time) and the Solar Radiation and Climate Experiment (SORCE). 
UARS covers wavelengths 22-420 nm. 
SORCE covers  wavelengths 0.5-2400 nm. 
SORCE began in 2003, after the 2000-2001 Cassini Jupiter flyby. 
We first average the UARS 22-420-nm data from October 2000 to March 2001. 
Then, we scale the 420-1000-nm SORCE data from 2003 to the time of the Cassini flyby by the SORCE to UARS ratio of integrated irradiance at 22-420 nm, combine the data from the two instruments, and scale for Jupiter's orbital distance.
The result is then modified to account for the Cassini camerasÕ system transmission and quantum efficiency \citep{west10}. 
We multiply the modified solar spectral irradiance by the area of Jupiter to get the reference disk-integrated  irradiance. 
The observed irradiance at each pixel is multiplied by the pixel area and summed over the pixels to get the observed disk-integrated  irradiance. 
The ratio between the observed and reference values is the full-disk albedo $p$, or disk-averaged $I/F$. } 
}}} 
\cora{The phase angle changes from 0{\deg} to $\sim$141{\deg} during the flyby, but there are no suitable full-disk images at some phase angles.
As a result, there are some gaps in the coverage, which are different for each filter. }

In addition to the Pioneer and Cassini data studied here, images of Jupiter were taken by Voyager, Galileo, and New Horizons, which covered a variety of angles. 
We are not aware of any other published data of the scattering phase functions for the Jovian surface.
Data for $\alpha>$150{\deg} do not exist because these directions would risk pointing spacecraft cameras too close to the Sun.
Accordingly, the fitted \corb{Henyey-Greenstein} curves at $\alpha>$150{\deg} are not well restricted. 
However the resulting light curves and spherical albedos are not severely affected by our extrapolation at $\alpha>$150\deg. 
At these angles the observed crescent is narrow, and the total reflected light is small.

\subsubsection{Saturn}
\label{sec: saturn}
For Saturn, we use the scattering phase function from \cite{dones93}, \corb{also used by} \cite{dyudina05}, which depends on three angles: incidence (via $\mu_0$), emission (via $\mu$), and phase angle $\alpha$.
The function is the Barkstrom law fitted by \cite{dones93} to Pioneer 11 \corb{modeling retrievals of \citep{tomasko84}} .
\begin{equation}
I/F={A \over \mu} \left({{\mu\cdot \mu_0}\over{\mu+\mu_0}}\right)^B~~,
\label{eq: backstorm}
\end{equation}
where 
$A$ and $B$ depend on the phase angle $\alpha$.
In this study we use the same functions as in \cite{dyudina05}. 
Table \ref{tab:backstorm} lists the Barkstrom parameters.
\begin{table}[htbp]
  \begin{center}
       \begin{tabular}{|c|c|c|c|c|c|c|c|c|}
\hline
Phase angle $\alpha$            &0\deg&30\deg&60\deg&90\deg &120\deg &150\deg &180\deg\\
\hline
A (red, 0.64 \mmdot)       &1.69   &1.59    &1.45  &1.34   &1.37   & 2.23  &3.09\\
B (red, 0.64 \mmdot)       &1.48   &1.48    &1.46  &1.42   &1.36   & 1.34  &1.31\\
A (blue, 0.44 \mmdot)      &0.63   &0.59    &0.56  &0.56   &1.69   & 1.86  &3.03\\
B (blue, 0.44 \mmdot)      &1.11   &1.11    &1.15  &1.18   &1.20   & 1.41  &1.63\\
\hline     
      \end{tabular}
    \caption{Coefficients for the Barkstrom function for Saturn, from \cite{dones93}, and also used by \cite{dyudina05}.}
    \label{tab:backstorm} 
  \end{center}
\end{table}

\subsection{Full-disk and Geometric Albedos}

To produce light curves of the fiducial exoplanets that we model, images of the planet for a set of locations along the orbit are generated using a model by \cite{dyudina05}.
For each image we integrate the total light coming from the planet to obtain the full-disk albedo $p(\alpha)$.
\begin{equation}
p(\alpha)={\sum_{\rm pixels} I(\mu,\mu_0,\alpha)
\cdot r_{\rm pix}^2/F \over \pi R_P^2}~~~,
\label{eq:pixel_integral}
\end{equation}
where $r_{\rm pix}$ is the pixel size and $R_P$ is the planet's radius.
Note that $ \sum_{\rm pixels} I(\mu,\mu_0,\alpha)\cdot r_{\rm pix}^2= L_P$ is the planet's luminosity.
Generally, $I$, $L_P$, $F$, and $p$ depend on wavelength.
In our case these values are weighted averages over the Pioneer and Cassini filter passbands \cora{(Fig. \ref{fig: cassini_fit}g for Cassini)}.
Geometric albedo $A_g$ (measured for extrasolar planets at secondary eclipse) is defined as our full-disk albedo at zero phase angle $\alpha=0$  (opposition), \ie, $A_g=p(0).$

%% file: results_include.tex


\section{Light Curves}
\label{sec:jupiter_saturn}

We first tested how light curves would differ for exoplanets with the surface reflection characteristics of Jupiter's and Saturn's, \corb{and with} Lambertian and Rayleigh scattering.

\begin{figure}[htbp]
\resizebox{!}{2.6in}{\includegraphics{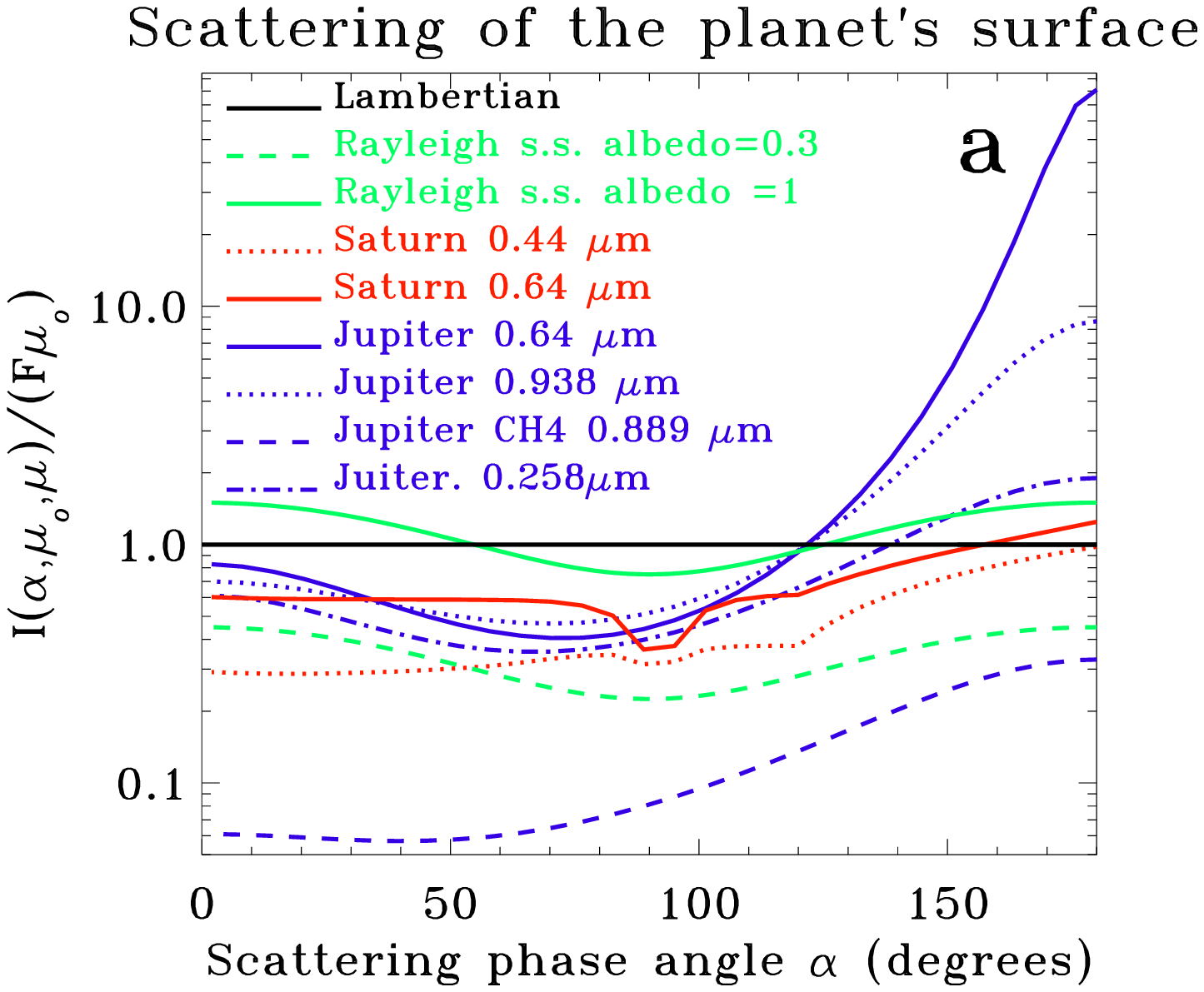}}
\hspace{-0.1in}
\resizebox{!}{2.6in}{\includegraphics{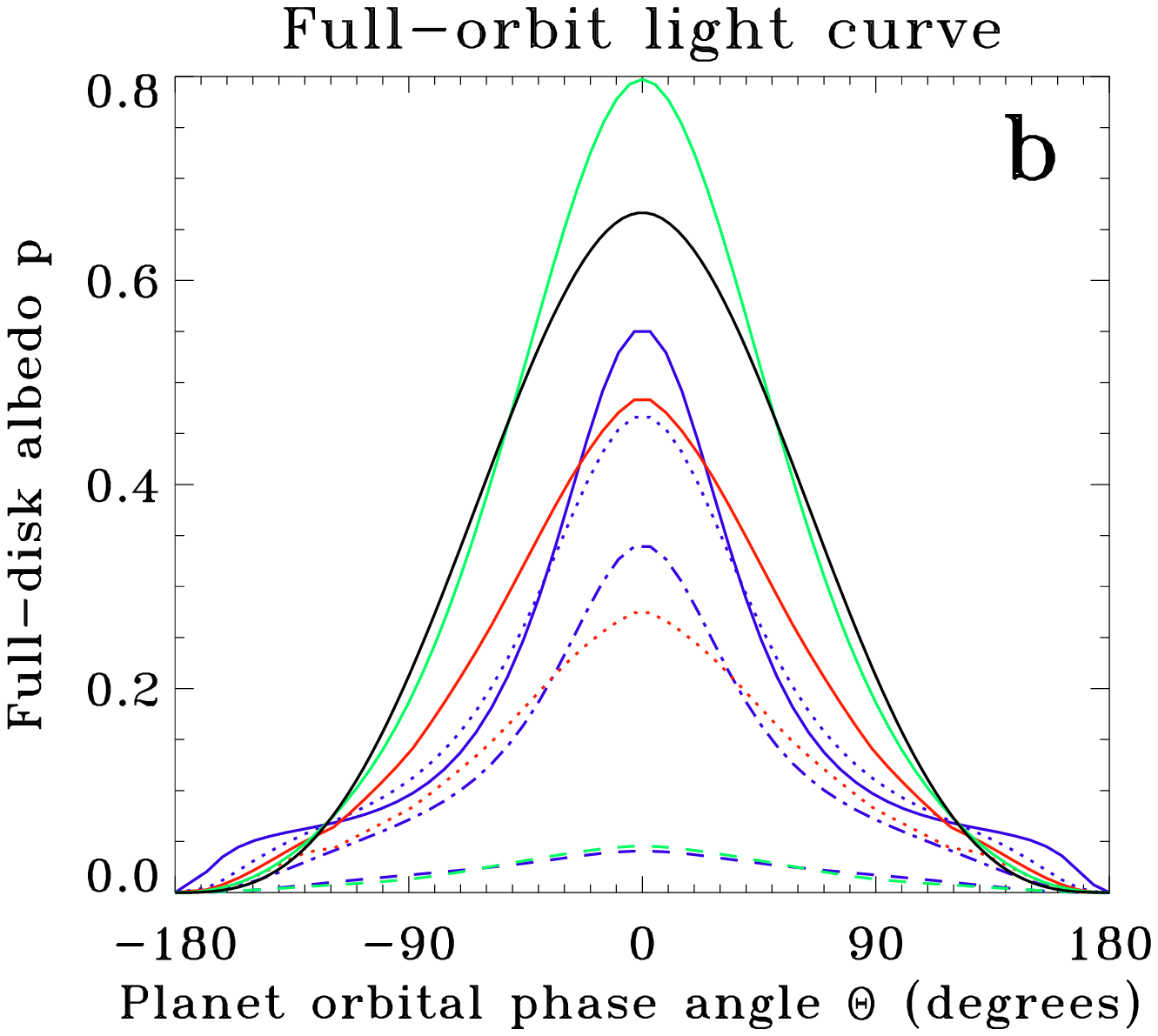}}
\newline
\resizebox{!}{2.6in}{\includegraphics{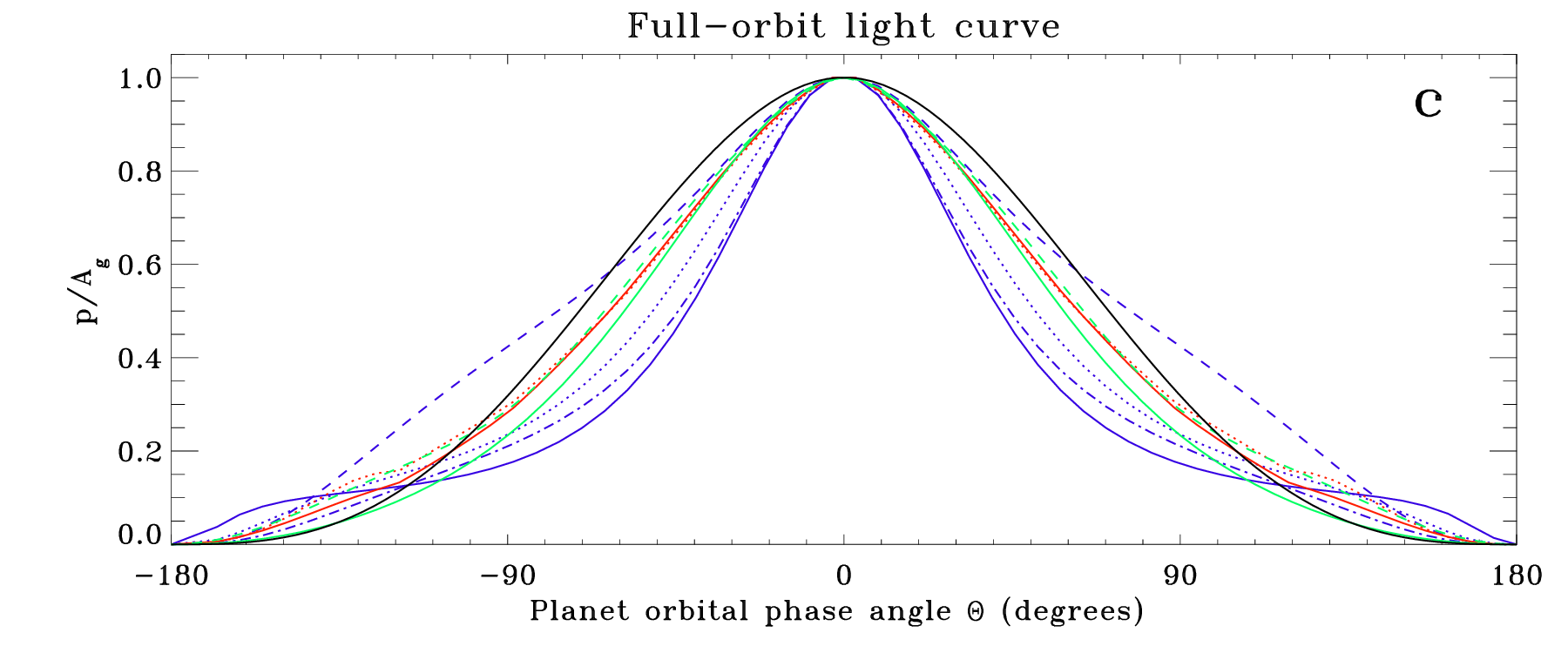}}

\caption{Panel a: Surface scattering functions for a Lambertian surface, semi-infinite Rayleigh layer with single scattering albedo 1 (green solid line)  
and single scattering albedo 0.3 (green dashed line), 
and Saturn and Jupiter at several wavelengths.
\cora{The linestyles are different from the notation in Section \ref{sec:model}. 
Here the linestyles represent wavelength bands.
For each wavelength only one curve labeled ``solid" in Section \ref{sec:model} is plotted in this figure as dashed, dotted or dot-dashed (see labels in panel a).}
The phase functions \cora{in panel a} are plotted for a specific geometry in which the Sun is 2\deg$ $  above the horizon ($\mu_0$=0.035) and the observer moves from the Sun's location ($\alpha=0$\deg) across the zenith toward the point on the horizon opposite to the Sun ($\alpha=178$\deg). 
Panel b: Comparison of light curves for a spherical planet assuming scattering properties from panel a.
The linestyles and colors correspond to panel a.
Panel c: \cora{Same as panel b but }all the curves are normalized by their geometric albedo $ A_g\equiv p(0)$.
\corb{The curves in digital form for panel b are available as a supplement.}
}
\label{fig:jup_sat}
\end{figure}

Figure \ref{fig:jup_sat}a \cora{compares} surface reflection functions 
for Jupiter and Saturn \cora{for the Pioneer and Cassini wavelength bands (fitted to the data in Figs. \ref{fig:tomasko78fit} and \ref{fig: cassini_fit}}), for a Lambertian surface, and for a Rayleigh-scattering surface. 
The Rayleigh case was calculated with the multiple scattering model VLIDORT \citep{spurr06} applied to a spherical planet, similarly to the results of \cite{kopparla15}, who used the quadrature method of \cite{horak50} for disk integration.
Note the logarithmic scale of the ordinate and the large amplitudes of the phase functions.

Figure \ref{fig:jup_sat}b compares edge-on light curves for spherical planets \corb{of the same radius} on circular orbits.
Their surface reflection properties correspond to panel a.
The luminosity of the planet is normalized by the incident stellar light to obtain the full-disk albedo $p$ as described in equation (\ref{eq:pixel_integral}).
The full-disk albedo can be converted into the planet's luminosity $L_P$ as a fraction of the star's luminosity $L_*$ for a planet of radius $R_P$ at an orbital distance $D_P$.
\begin{equation}\label{eq:lp_l*}
L_P/L_*=(R_P/D_P)^2\cdot p
\end{equation}
For example, for Saturn at 1 AU, 
$(R_P/D_P)^2\approx$1.6\tms$10^{-7}$.
The plot in Fig. \ref{fig:jup_sat}b can be transformed into a time-dependent light curve simply by dividing orbital phase angle $\alpha$ by 360\deg\ and multiplying by the planet's orbital period.

The variety of light curve shapes in Fig. \ref{fig:jup_sat}c \corb{demonstrates  the uncertainty range in realistic cloudy atmospheres}. 
For example, \cora{some of Jupiter's light curves \corb{in Fig. \ref{fig:jup_sat}c} would be a factor of two} fainter at half phase ($\alpha\approx\pm 90$\deg) than the Lambertian curve, given the same geometric albedos at secondary eclipse ($\alpha= 0$\deg).
Also, near the transit ($\alpha\approx\pm 180$\deg), the Lambertian curve \corb{strongly underestimates} the more realistic light curves\corb{, which are brighter due to forward scattering}.
\corb{A similar forward scattering effect is also seen in light curves of Mars, Venus, and Jupiter \citep{sudarsky05, dyudina05}, which is essential for characterizing extrasolar planets from light curve observations near transit.}

As can be seen in Fig. \ref{fig:jup_sat}b, 
the light curves for Jupiter peak more sharply at \cora{opposition ($\Theta\approx$ 0\deg)} than the light curves of Saturn, a Lambertian planet, or a Rayleigh-scattering planet. 
This \cora{reflects} the sharp back-scattering peak in Jupiter's \cora{surface} scattering  (at $\alpha\approx$ 0\deg$ $ in Fig. \ref{fig:jup_sat}a).
The peak is probably the result of scattering by large cloud particles.
A Rayleigh scattering atmosphere 
 (green curves in Fig. \ref{fig:jup_sat}) can also produce \corb{sharper backscattering peaks than the} Lambertian case
 , as previously modeled by \cite{madhusudhan12}.
As seen in Fig. \ref{fig:jup_sat}c, the peaks in the light curves for Jupiter (blue lines) are much sharper than the ones expected from Rayleigh scattering (green lines).
\corb{In this paper we explore the variation of  the backscattering peak with wavelength for Jupiter.}

\begin{figure}[htbp]
\resizebox{!}{2.6in}{\includegraphics{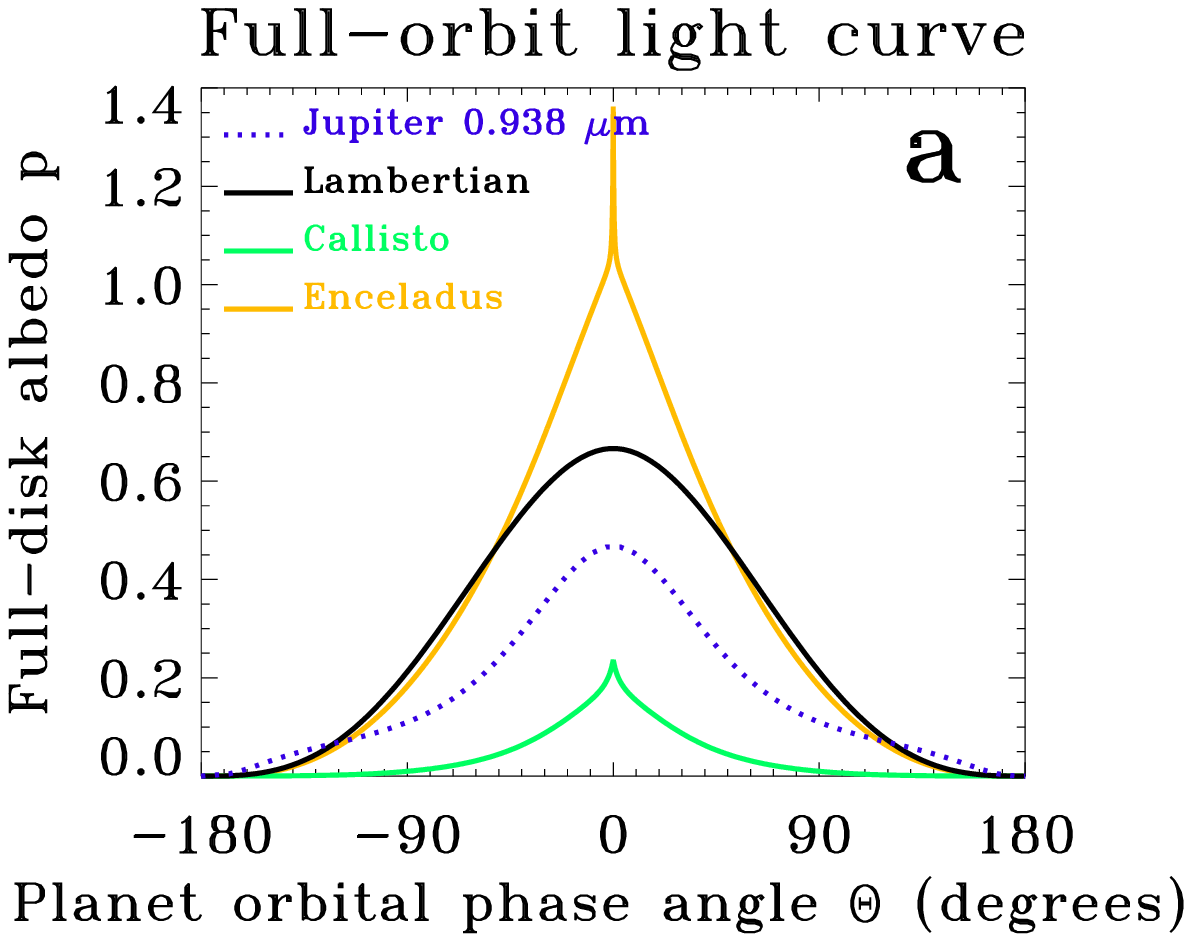}}
\resizebox{!}{2.6in}{\includegraphics{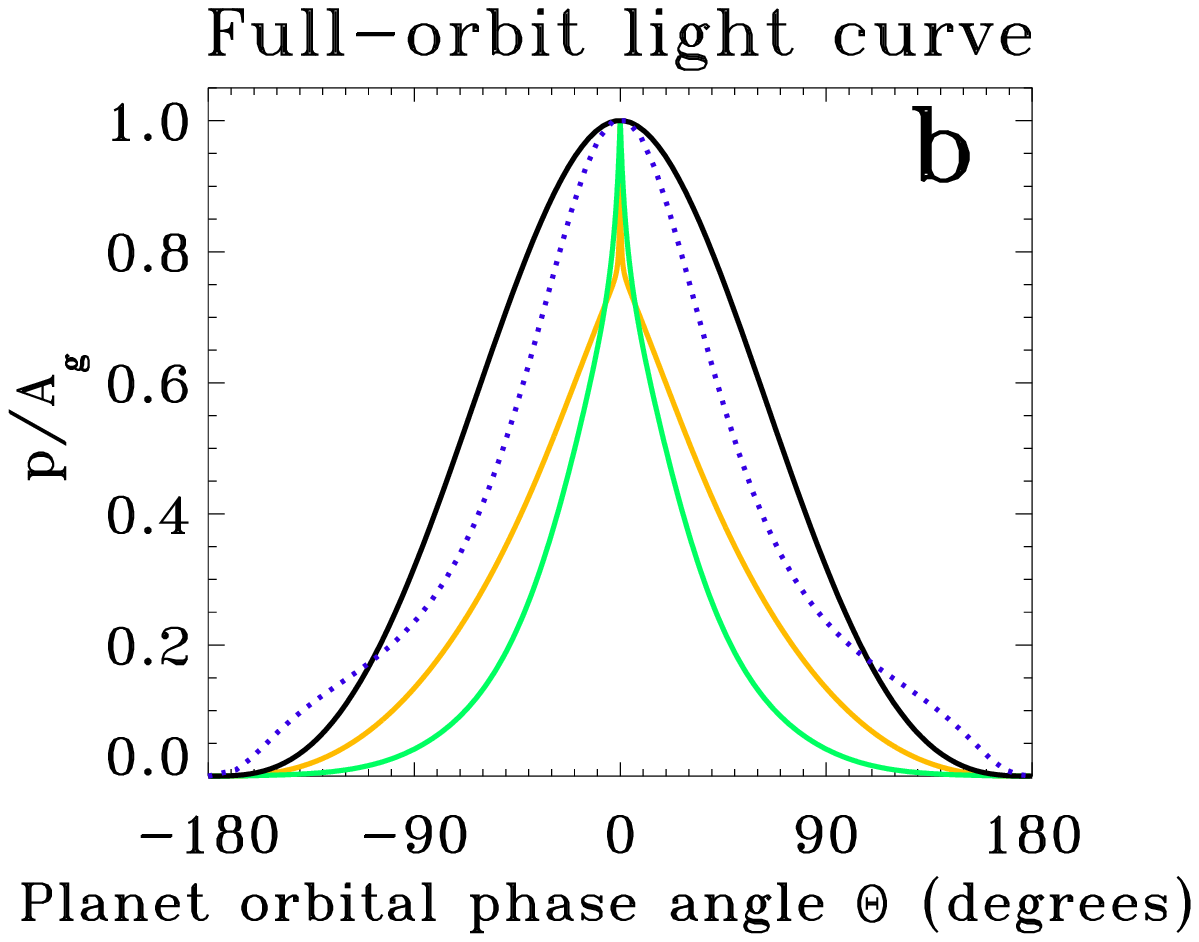}}
\caption{\corb{Panel a: Comparison of our atmospheric window light curve for Jupiter with the light curves of Solar System moons.
The Callisto light curveat at 0.47\mm was derived from Voyager clear filter and groundbased observations \citep{domingue97} 
}. 
\corb{The Enceladus light curve is combined from Hubble Space Telescope F439W filter observations (434 \mmdot) and Voyager clear filter data \citep{verbiscer05, verbiscer07}. 
Panel b: Same as panel a but all the curves are normalized by their geometric albedo $A_g$. }
\corb{The satellite curves in digital form for panel a are available as a supplement.}}
\label{fig: moons}
\end{figure}

\corb{Backscattering peaks were previously observed in the phase functions on the Moon, Mars, Uranus, Venus and Jupiter \cite[see summary in Fig. 3 of ][]{sudarsky05}.}
\corb{The atmospheres of Uranus and Venus produce backscattering peaks similar to the ones we see for Jupiter, while solid surfaces such as the Moon and Mars produce sharper peaks. 
Solar System moons without atmospheres exhibit sharp backscattering peaks that are the products of contributions from the opposition effect{\footnotemark}
\footnotetext{\corb{The opposition effect is the dramatic, non-linear increase in reflectance seen near opposition.  It is produced by two effects: shadow hiding and coherent backscattering. The shadow hiding opposition effect takes place from phase angles 0-20 deg; the coherent backscattering opposition effect takes place at much smaller phase angles, from 0-2 deg.}} 
and the single particle phase function \citep{domingue97,deau09}.
Figure \ref{fig: moons} compares our light curve for Jupiter in an atmospheric window with the light curves of Saturn's high-albedo moon Enceladus and Jupiter's low-albedo moon Callisto{\footnotemark}.}
\footnotetext{\corb{
The phase curve of Callisto from \cite{domingue97}, is derived from fits to the \cite{hapke93b} photometric model of full-disk Voyager clear filter (0.47  \mmdot) and ground based observations (G.W. Lockwood and D.T. Thompson).
The curve is for the trailing hemisphere of Callisto since the phase angle coverage is best for that hemisphere. 
It is generated using the Hapke parameters  \citep[Table II]{domingue97}. }}
\corb{The backscattering peaks for the moons are very narrow -- they span, at most, a few degrees from opposition.
Such peaks are due to the coherent backscatter opposition effect \citep{hapke93, shkuratov94}, a sharp increase in reflectance produced by the constructive interference of incident and reflected light rays. 
Similar very narrow opposition peaks are modeled for cold cloudy giant exoplanets \citep[][ Fig. 4]{sudarsky05}.
Cassini observations of Jupiter used in this work do not cover angles that close to opposition  (Cassini data are for $\alpha>3.4$\deg).
Accordingly, our curves may underestimate the strength of the opposition peak on Jupiter at low phases ($\alpha<3.4$\deg).
More (possibly Earth-based) full-disk observations or additional analysis of partial-disk Cassini images near opposition can provide a better restriction on Jupiter's backscattering peak.}



%
\section{Total Reflection from the Planet}
\label{sec: spherical_albedo}


The total reflected light from the planet can be obtained by integrating the reflected light in all directions and wavelengths.
The result divided by the incident solar flux is called the Bond albedo $A_b$.
$A_b$ is usually estimated  from the observable geometric albedo $A_g$ at a particular wavelength band,  extrapolated over wavelengths and over different directions.
The Lambertian assumption is commonly used, which, as we show here, is likely to overestimate the Bond albedo.

At a specific wavelength, the reflection of the planet's surface 
is defined by its spherical albedo $A_S$.
\begin{equation}
A_S=\int_0^{4\pi}p(\alpha) d \Omega,
\label{eq: a_s}
\end{equation}
where the reflection from the planet is integrated over all outgoing solid angles $\Omega$.
In the general case $p$ also depends on the azimuth of the observer relative to the planet.
In our simplified case $p(\alpha)$ depends only on phase angle $\alpha$, and $A_S$ can be converted to an integral over $\alpha$ (see also Eq. 3.29 of \cite{seager10b}, where $\Psi=\pi \cdot p(\alpha)$).
\begin{equation}
A_S=2\int_0^{\pi}p(\alpha) \sin (\alpha) d\alpha,
\label{eq: a_s_alpha}
\end{equation}
\cora{The absorbed light for the planet, as a fraction of incident light, is $1-A_S$.}
In the Lambertian case the integral over $\alpha$ can be solved analytically, giving the theoretical ratio of \cora{spherical to geometric albedo, the phase intergal $A_S/A_g$}:
\begin{equation}
A_S={3\over2} p(\alpha=0) = {3\over2} A_g
\label{eq: a_s_lambert}
\end{equation}

We calculate spherical albedos in the observed spacecraft filter bands to estimate how the Lambertian approximation can over- or under-estimate the conversion from geometric to Bond albedo, which is typically used to estimate the stellar heating for extrasolar planets.
Table \ref{tab: albedo} lists the geometric and spherical albedos derived in our \cora{models}.  
\begin{table}[htbp]
\caption{Geometric and spherical albedos \cora{$A_g$ and $A_S$, and the phase integral $A_S/A_g$ derived for different \cora{reflection functions fitted to the data}.
The fitting uncertainty is indicated by the range of the curves for each filter (solid, dashed, and dotted in Figs. \ref{fig:tomasko78fit} and \ref{fig: cassini_fit} and Table \ref{tab: hg}).
\corb{The albedos for the moons Callisto and Enceladus are derived from their light curves (Fig. \ref{fig: moons}).} 
The coefficient for the non-Lambertian correction C is given in the last column.
}}
\vskip4mm
\centering
\begin{tabular}{|c|c|c|c|c|c|c|c|}
\hline  
 & Geometric & Spherical  & $A_S/A_g$ & C \\
 & albedo  $A_g$ &   albedo $A_S$    &   & \\
\hline
Lambertian&0.667&1.00& 1.50&1.00 \\
\hline
Rayleigh semi-infinite s.s. albedo = 1&0.676&0.94& 1.40&0.93 \\
\hline
Saturn 0.44 (0.39-0.5)$\mu$m  (Pioneer blue)&0.274&0.39& 1.44&0.96 \\
\hline
Saturn 0.64 (0.59-0.72)$\mu$m  (Pioneer red)&0.483&0.67& 1.39&0.92 \\
\hline
Jupiter 0.64 (0.59-0.72)$\mu$m (Pioneer red) \cora{solid}&0.549&0.56& 1.02&0.68 \\
\cora{dashed} &0.508&0.48& 0.95&0.63 \\
\cora{dotted} &0.560&0.71& 1.28&0.85 \\
\hline
Jupiter atm. window 0.938 $\mu$m (Cassini CB3) \cora{solid}&0.466&0.57& 1.23&0.82 \\
\cora{dashed} &0.474&0.48& 1.01&0.67 \\
\cora{dotted} &0.456&0.74& 1.63&1.08 \\
\hline
Jupiter CH$_4$ 0.889 $\mu$m (Cassini MT3) \cora{solid}&0.040&0.07& 1.72&1.14 \\
\cora{dashed}&0.043&0.07& 1.62&1.08 \\
\cora{dotted}&0.040&0.06& 1.55&1.03 \\
\hline
Jupiter 0.24-0.28 $\mu$m (Cassini UV1) \cora{solid}&0.339&0.36& 1.07&0.71 \\
\cora{dashed}&0.364&0.27& 0.74&0.49 \\
\cora{dotted}&0.310&0.41& 1.32&0.88 \\
\hline
\corb{Callisto}&0.235&0.09& 0.40&0.26 \\
\hline
\corb{Enceladus}&1.362&1.0 & 0.74&0.49 \\
\hline     
      \end{tabular}
    \label{tab: albedo}
\end{table}
We also list the phase integral $A_S/A_g$, \cora{to compare it with previous results}.
The phase integral  had been estimated before from Pioneer images at red and blue wavelengths. 
A cloud scattering model  fitted to Pioneer data gives $A_S/A_g$=1.2-1.3  \cora{for Jupiter} \citep{tomasko78}, and  $A_S/A_g$=1.4$\pm0.1$ for Saturn \citep{tomasko84}.
Our estimates, based on curve fitting to the surface scattering data \corb{(Saturn and Jupiter atmospheric window band in Table \ref{tab: albedo})}, \cora{are consistent with these values}.

\corb{Modeled phase integrals for extrasolar planets \citep[][Fig. 8]{ sudarsky05} range between 0.25 to 1.5 for wavelengths below 1 $\mu$m.
The extremely low phase integrals ($A_S/A_g$=0.3-0.5) in methane absorption bands derived in that model are not seen in our retrieval.
Instead the phase integral in our 0.889 \mm methane absorption band ranges $A_S/A_g$=1.55-1.72, which is somewhat higher than in the 0.938 \mm atmospheric window ($A_S/A_g$=1.01-1.63).
This is probably determined by the clouds and hazes on Jupiter.}

The phase integrals in Table \ref{tab: albedo} are useful for understanding the realistic conversion from geometric to  Bond albedo $A_b/A_g$.
Without spectral information, $A_b$ can be estimated from the geometric albedo measured at a particular wavelength assuming $A_S$ is constant with wavelength.
Then the Bond albedo, averaged over wavelength, \cora{can be roughly approximated by spherical albedo}: $A_b\sim A_S$.

Looking at the phase integral  $A_S/A_g$ at different wavelength in Table \ref{tab: albedo}, one can get a feel for the range of possible $A_b/A_g$ ratios for Saturn- and Jupiter-like atmospheres.
Because the common assumption for exoplanets is Lambertian ($A_S/A_g=3/2$), we calculate a correction coefficient $C$.
It shows an overestimate which the Lambertian assumption imposes on the total planet's reflection when it is derived from geometric albedo. 
 
\begin{equation}
C={(A_S/ A_g)\over(A_S/ A_g)_{Lambertian}}={2 (A_S/ A_g)\over 3 }
\end{equation}

To correct the spherical albedo derived under the Lambertian assumption  $A_S(Lambertian)$ for realistic anisotropic scattering, it should be multiplied by $C$: $A_S=C\cdot A_S(Lambertian)$.
\corb{For the} wavelength-averaged Bond albedo, \corb{such correction} is probably in the range of coefficients \corb{listed for} different wavelength bands in Table \ref{tab: albedo}.
Table \ref{tab: albedo} shows that it may be as low as $C=$0.68, which means that Lambertian assumption gives an overestimate for $A_S$ by a factor of $1/0.68\approx1.5$.

To put it in the context of exoplanets, for the planet HD 189733b, having $A_g\approx0.4$ at blue wavelengths \citep{evans13},  would mean that instead of 40\% of stellar light ($1-A_S(Lambertian)=1-A_g\times(A_S/ A_g)_{Lambertian}=1-0.4\times1.5=40\%$), the planet absorbs about 60\% ($1-A_S(Lambertian)\times0.68\approx60\%$). 


%% file: discussion_include.tex
\section{Discussion}
\label{sec: discussion}

This research presents a summary of Jupiter's and Saturn's cloud reflection properties at different wavelengths and illumination phases.
\corb{We fitted} simple analytical functions to the observations.
These summarized observations \corb{do not address the physics of cloud formation.
To derive a cloud distribution from these data, light scattering models are needed. 
However, the simplified interpolated functions derived here can be easily used 
to test planet-averaged extrasolar cloud models using Jupiter and Saturn as examples.
For Jupiter and Saturn, our planet-averaged functions are oversimplified.
Spatially-resolved cloud reflection data allow one to derive the cloud distribution more accurately with the help of radiative transfer and cloud microphysics models of Jupiter and Saturn.} 

We derived light curves consistent with Jupiter's and Saturn's spacecraft observations.
To do that we used the direct measurements from full-disk spacecraft images, and, where such images are not available, extrapolation of this data with a disk-averaging model that converts partial-disk images to full-disk.

The light curves for the cloudy atmospheres of Jupiter and Saturn (Fig. \ref{fig:jup_sat}) show considerably different shapes than the Lambertian light curve. 
\corb{For extrasolar planets it means that a factor of few differences in amplitude are expected at phases other than secondary eclipse ($\Theta=0$)}. 
Especially interesting 
is the high brightness due to forward scattering near the transit \corb{($\Theta=\pm180$)}, which is not present in the Lambertian case. 
However, this forward-scattering effect is not well restricted by the data on Jupiter and Saturn because the spacecraft avoid pointing too close to the Sun  to prevent the detectors from overheating.

The total reflected flux from the planet judged by geometric albedo also depends on forward and backward scattering.
From the variety of wavelength bands studied (Table \ref{tab: albedo}), the effect is between a factor of \cora{0.6 and 1.2} (values of C in Table \ref{tab: albedo}).

\corb{Exoplanets span a variety of both compositions and insolation fluxes. 
Light curves derived here for Jupiter and Saturn are relevant to high-albedo cloudy exoplanets.
For example, such planets were modeled by \cite{sudarsky05}: 1M$_J$, 5 Gyr planets at a G2 V star orbiting at distances larger than 2 AU.
Our light curve shapes for the atmospheric window (CB3) and methane absorption band (MT3) are probably representative for atmospheric windows and absorption bands produced by other gases on planets with bright condensate clouds, regardless of cloud composition.
The ultraviolet broadband UV1 channel is sensitive to specific photochemical hazes typical for Jupiter, and the corresponding light curve may only be applicable to Jupiter analogues around Sun-like stars.
Our work may also provide insights into some cooler super Earths \citep{morley13}.
}

\corb{Solid bodies have much stronger backscattering at opposition than cloudy planets, as seen on example of Solar System moons in Fig. \ref{fig: moons}.
For extrasolar planets opposition effect on the light curve near secondary eclipse is likely to be a good indicator of whether the planet is solid or gaseous.
It should be noted that the solid-body backscattering peak is so narrow that the angular size of the star as seen from such planet 
will widen the peak, especially for close-in planets.
For solid planets opposition effect would result in much lower Bond albedos and stronger heating than expected from geometric albedo observations}.

%% file: supplementary_captions.tex
\section*{Online Supplementary Data Captions}



\corb{The observational data from Fig. 2, 3, and 4 are given in the text format as separate files for each data set.
The file names and text headers within each file indicate the details of each data set.}

{\bf Numerical data for Figure 2.}

\ni\corb{Files fig\_2\_a\_Cassini\_pixel\_brightness\_CB3.txt,} 

\ni\corb{fig\_2\_c\_Cassini\_pixel\_brightness\_MT3.txt,} 

\ni\corb{and fig\_2\_e\_Cassini\_pixel\_brightness\_UV1.txt list pixel values of wide angle camera (WAC) images taken with a particular filter during Cassini Jupiter flyby.
The pixel brightness is listed in I/F units.
Also, for each pixel, the following values are listed: cosine emission angle $\mu$, cosine incidence angle $\mu_0$, phase angle $\alpha$ in degrees,  planetocentric  latitude, and   longitude in degrees.
The data processing is done using standard calibration \citep{west10} and described in \cite{zhang13}}

\corb{Files fig\_2\_b\_full\_disk\_Cassini\_CB3.txt, fig\_2\_d\_full\_disk\_Cassini\_MT3.txt, and fig\_2\_f\_full\_disk\_Cassini\_UV1.txt contain full-disk albelo $p$ derived from WAC images taken with a  particular filter  during Cassini Jupiter flyby. 
The albedos are listed depending on the phase angle  $\alpha$ in degrees.
The data are processed with standard calibration \citep{west10} in the group of co-author Li.
}

{\bf Numerical data for Figure 3.}

\corb{Files fig\_3\_b\_lambertian. txt, fig\_3\_b\_jupiter\_CB3.txt É etc. contain the 9 light curves simulated in this paper, which are shown in Fig. 3 b.
Filter names correspond to Fig. 3 labels as follows: CB3 is 0.938\mmdot, MT3 is 0.889\mmdot, UV1 is 0.258\mmdot, red is 0.64\mmdot, blue is 0.44\mmdot.
The full-disk albedo $p$ is listed versus planet's orbital phase angle (in degrees).}

{\bf Numerical data for Figure 4.}

\corb{Files  fig\_4\_a\_Enceladus.txt and fig\_4\_a\_Callisto.txt contain light curves for Enceladus  and Callisto, which are shown in Fig. 4a }